\documentclass[twoside]{IEEEtran}
\usepackage{amssymb}
\usepackage{graphicx}

\begin{document}
\newtheorem{Theorem}{Theorem}
\newtheorem{Prop}{Proposition}
\newtheorem{Coro}{Corollary}
\newtheorem{Lemma}{Lemma}

\title{Catalyst-assisted Probabilistic Entanglement Transformation}

\author{Yuan Feng,\ Runyao Duan
and Mingsheng Ying\thanks{This work was partly supported by the
National Foundation of Natural Sciences of China (Grant Nos:
60496321, 60321002, and 60433050) and by the Key grant Project of
Chinese Ministry of Education (Grant No: 10402).}\thanks{ The
authors are with the State Key Laboratory of Intelligent
Technology and Systems, Department of Computer Science and
Technology, Tsinghua University, Beijing, China 100084. E-mails:
feng-y@tsinghua.edu.cn (Yuan Feng), dry02@mails.tsinghua.edu.cn
(Runyao Duan) and yingmsh@tsinghua.edu.cn (Mingsheng Ying)}}

\date{\today}
\maketitle
\begin{abstract}
We are concerned with catalyst-assisted probabilistic entanglement
transformations. A necessary and sufficient condition is presented
under which there exist partial catalysts that can increase the
maximal transforming probability of a given entanglement
transformation. We also design an algorithm which leads to an
efficient method for finding the most economical partial catalysts
with minimal dimension. The mathematical structure of
catalyst-assisted probabilistic transformation is carefully
investigated.
\end{abstract}

\section{Introduction}

\PARstart{Q}{uantum} entanglement plays an essential role in
quantum information processing \cite{NC00}. Indeed, it is a
necessary resource in quantum cryptography \cite{BB84}, quantum
superdense coding \cite{BS92}, and quantum teleportation
\cite{BBC+93}, which are striking tasks in quantum information
processing. When entanglement is treated as a type of resource,
the study of how to quantify and manipulate it becomes crucial
(for a survey of quantum information theory, we refer to
\cite{BS98}). A fruitful research direction is to try to discover
the laws that must be obeyed when transforming between different
forms of entanglement using only local operations on the separate
subsystems and classical communication between them. This kind of
transformation is usually abbreviated as LOCC. The communication
constraints that characterize LOCC are fundamentally and
practically important, since many applications of quantum
information processing involve spatially separated parties who
must manipulate an entanglement state without performing joint
operations.

Suppose two spatially separated parties, Alice and Bob, share a
bipartite quantum state $|\psi_1\rangle\in \mathcal{C}^n \otimes
\mathcal{C}^n$ with Schmidt decomposition
$$|\psi_1\rangle=\sum_{i=1}^n \sqrt{\alpha_i}|i_A\rangle |i_B\rangle,$$
where $\alpha_1\geq \dots\geq \alpha_n\geq 0$ are the Schmidt
coefficients of $|\psi_1\rangle$ and $\sum_i\alpha_i=1$.
$|i_A\rangle$ and $|i_B\rangle$ are orthonormal bases of Alice's
and Bob's systems, respectively. Suppose the parties want to
transform this initial state into a desired bipartite state
$|\psi_2\rangle$ with Schmidt decomposition
$$|\psi_2\rangle=\sum_{i=1}^n \sqrt{\beta_i}|i'_A\rangle |i'_B\rangle,$$
where $\beta_1\geq \dots\geq \beta_n\geq 0$, and
$\sum_i\beta_i=1$. The orthonormal bases $|i'_A\rangle$ and
$|i_A\rangle$ (also $|i'_B\rangle$ and $|i_B\rangle$) are not
necessarily the same. Nielsen \cite{NI99} proved that Alice and
Bob can realize this transformation from $|\psi_1\rangle$ to
$|\psi_2\rangle$ by LOCC if and only if
$$\sum_{i=1}^l \alpha_i \leq \sum_{i=1}^l \beta_i {\rm \ \ \ for\ any\ \ \ }1\leq l\leq n,$$
with equality holding when $l=n$, or equivalently, by the theory
of majorization \cite{MO79}\cite{AU82}, $\lambda_{\psi_1}$ is
majorized by $\lambda_{\psi_2}$, written
$$\lambda_{\psi_1}\prec\lambda_{\psi_2},$$ where the probability
vectors $\lambda_{\psi_1}$ and $\lambda_{\psi_2}$ denote the
Schmidt coefficient vectors of $|\psi_1\rangle$ and
$|\psi_2\rangle$, respectively.

Nielsen's work establishes a connection between the theory of
majorization in linear algebra  and entanglement transformation.
Furthermore, since the necessary and sufficient condition
mentioned above is very easy to check, it is extremely useful in
telling whether one bipartite entangled state can be transformed
into another by LOCC. Nielsen's theorem directly implies that
there exist incomparable states in the sense that any one cannot
be transformed into another only using LOCC. To treat the case of
transformations between incomparable states, Vidal \cite{Vidal99}
generalized Nielsen's work by allowing probabilistic
transformations. He found that although deterministic
transformation cannot be realized between incomparable bipartite
pure states, a probabilistic one is always possible (notice that
when multipartite states are considered, this statement does not
hold \cite{DV00}). Furthermore, he gave an explicit expression of
the maximal probability of transforming one state to another. To
be more specific, let $P(|\psi_1\rangle \rightarrow
|\psi_2\rangle)$ denote the maximal probability of transforming
$|\psi_1\rangle$ into $|\psi_2\rangle$ by LOCC. Then
$$P(|\psi_1\rangle \rightarrow |\psi_2\rangle)=\min_{1\leq
l\leq n} \frac{\sum_{i=l}^n \alpha_i}{\sum_{i=l}^n
\beta_i}=\min_{1\leq l\leq n}
\frac{E_l(\lambda_{\psi_1})}{E_l(\lambda_{\psi_2})},$$ where
$E_l(\lambda_{\psi_1})$ denotes $\sum_{i=l}^n \alpha_i$. In what
follows, we extend this notation to any probability vector.

Another interesting phenomenon discovered by Jonathan and Plenio
\cite{JP99} is that sometimes an entangled state can enable
otherwise impossible entanglement transformations without being
consumed at all. A simple but well known example is
$|\psi_1\rangle\nrightarrow |\psi_2\rangle$ but
$|\psi_1\rangle\otimes|\phi\rangle \rightarrow
|\psi_2\rangle\otimes|\phi\rangle$, where
$$|\psi_1\rangle=\sqrt{0.4}|00\rangle+\sqrt{0.4}|11\rangle+\sqrt{0.1}|22\rangle+\sqrt{0.1}|33\rangle,$$
$$|\psi_2\rangle=\sqrt{0.5}|00\rangle+\sqrt{0.25}|11\rangle+\sqrt{0.25}|22\rangle,$$
and
$$|\phi\rangle=\sqrt{0.6}|44\rangle+\sqrt{0.4}|55\rangle.$$
The role of the state $|\phi\rangle$ in this transformation is
analogous to that of a catalyst in a chemical process. The
mathematical structure of this phenomenon, so called
catalyst-assisted entanglement transformation, was carefully
examined by Daftuar and Klimesh \cite{DK01}. They found that there
does not exist an upper bound on the dimension of catalysts that
should be considered, in trying to determine which states can be
transformed into a given state. Furthermore, they proved that any
nonuniform state, which has at least two nonzero Schmidt
coefficients nonequal, can serve as a catalyst for some
entanglement transformation. On the other hand, Eisert and Wilkens
found that catalysis is also helpful in entanglement
transformations for bipartite mixed states \cite{EW00}.

In this paper, we examine the power of catalysis in probabilistic
entanglement transformations. We have noticed that in \cite{JP99}
Jonathan and Plenio found that in some cases, an appropriately
chosen catalyst can increase the maximal transformation
probability of incomparable states. The example they presented is
as follows. Let
$$|\psi_1\rangle=\sqrt{0.6}|00\rangle+\sqrt{0.2}|11\rangle+\sqrt{0.2}|22\rangle$$
and
$$|\psi_2\rangle=\sqrt{0.5}|00\rangle+\sqrt{0.4}|11\rangle+\sqrt{0.1}|22\rangle.$$
The maximal probability of transforming $|\psi_1\rangle$ into
$|\psi_2\rangle$ under LOCC is $80\%$ while when a catalyst
$$|\phi\rangle=\sqrt{0.65}|33\rangle+\sqrt{0.35}|55\rangle$$
is introduced, the probability can be increased to $90.4\%$. They
also showed that enhancement of the transformation probability is
not always possible. However, no further results on this topic
were given in their paper.

The main aim of our paper is to study the structure of catalysis
as applied to probabilistic entanglement transformations. We give
a necessary and sufficient condition for the existence of partial
catalysts (quantum states which can increase the maximal
transforming probability while not being consumed) for a given
entanglement transformation. Rather surprisingly, we find that
whether or not the probability can be increased depends only on
the minimal Schmidt coefficients of the original state and the
target state, provided that the maximal transforming probability
is less than 1 (the probability cannot of course be increased if
equal to 1). To be specific, a probabilistic transformation has
partial catalysts if and only if the maximal transforming
probability is less than the minimum of 1 and the ratio of the
minimal Schmidt coefficient of the original state to that of the
target state. Furthermore, we show that if the maximal probability
of a transformation can be increased by some catalyst, then there
is a sequence of $2\times 2$ dimensional states that increases the
maximal probability of the transformation.

For any given entanglement transformation, we present a systematic
way to construct partial catalysts. The catalysts are, however,
not economical in general in the sense that they do not
necessarily have the minimal dimension among all partial catalysts
for this transformation. In fact, the problem of constructing
systematically the most economical partial catalysts for any given
transformation seems to be hard and remains open. We can, however,
give a numerical solution to this problem by solving a series of
inequalities.

This paper is organized as follows. In Section II, we consider
briefly probabilistic entanglement transformations without the aid
of catalysis. We first provide a simple connection between
probabilistic transformations and deterministic ones, which is
helpful in realizing probabilistic transformations since
deterministic ones have been well researched. We then examine
properties such as monotonicity and continuity of the set of
states that can be transformed into a given state by LOCC with a
probability not less than a given positive number. Section III is
the main body of this paper. We present here a necessary and
sufficient condition under which a given probabilistic
transformation has partial catalysts. Moreover, the catalysts are
systematically constructed. To find the most economical ones, we
present first an algorithm to decide whether there exist partial
catalysts with a given dimension and find out all suitable ones.
Based on this algorithm, the mission of constructing the most
economical partial catalysts is achieved by applying the algorithm
to state spaces with increasing dimension from 2 (an upper bound
on the dimension we should consider can be predetermined because
we have constructed a non-economical one). In Section IV, we
generalize the result of Daftuar and Klimesh \cite{DK01} to the
set of states that can be transformed into a given state by
catalyst-assisted LOCC, with a probability not less than a given
positive number. We find that this set shares many properties with
the well known set which consists of all states being trumped by
the given state (for the latter set, we refer to \cite{DK01} for
details). To be more specific, the generalized set is convex and
not closed in general; the dimensions of catalysts that should be
considered in trying to determine the states in the set have no
upper bound. We further investigate the mathematical structure of
this generalized set and find out all the boundary and extreme
points. This gives an answer to Nielsen's open problem in the case
of deterministic transformations. Finally, a conclusion is drawn
and some open problems are presented in Section V.

For simplicity, in what follows we denote a bipartite quantum
state by the probability vector of its Schmidt coefficients. This
will not cause any confusion because it is well known that the
fundamental properties of a bipartite state under LOCC are
completely determined by its Schmidt coefficients. Therefore, from
now on, we consider only probability vectors instead of bipartite
quantum states and always identify a probability vector with the
corresponding quantum state. Sometimes we even omit the
normalization of positive vectors to be probability ones for the
sake of simplicity.

\section {Probabilistic Entanglement Transformation}

In this section, we discuss the structure of probabilistic
entanglement transformations without catalysis. Denote by $V^n$
the set of all $n$-dimensional probability vectors and let
$x,y,\dots$ range over $V^n$. Given a positive number $\lambda\leq
1$, let
$$S^\lambda(y)=\{x\in V^n:P(x\rightarrow y)\geq \lambda\},$$ which is the
set of $n$-dimensional probability vectors that can be transformed
into $y$ by LOCC with the maximal probability not less than
$\lambda$. When $\lambda=1$, the set reduces to the well known set
$S(y)$ which includes exactly the vectors that can be transformed
into $y$ with certainty, or equivalently, that are majorized by
$y$ (see \cite{MAJ} for details of $S(y)$).

What we would like to point out first is that there is a simple
relationship between probabilistic entanglement transformation and
the theory of weak majorization, just like the connection between
deterministic transformation and the theory of majorization
discovered by Nielsen in \cite{NI99}. Recall that an
$n$-dimensional positive vector $u$ is called super-majorized
\cite{MO79} by another $n$-dimensional positive vector $v$,
written $u\prec^\omega v$, if and only if
$$\sum_{i=l}^n u^\downarrow_i \geq \sum_{i=l}^n v^\downarrow_i$$
for each $l$ in the range 1 through $n$. Here $u^\downarrow$
denotes the vector obtained by arranging the components of $u$ in
nonincreasing order. Notice that the only difference between
super-majorization and majorization is the omission of the
equality requirement at $l=1$. It is very easy to check by G.
Vidal's formula that $x\in S^\lambda(y)$, or equivalently,
$P(x\rightarrow y)\geq \lambda$ if and only if the
super-majorization relation
$$x\prec^\omega \lambda y$$
 holds.

It is well known that there is a close connection between
majorization and doubly stochastic matrices \cite{HLP52}. To be
specific, for all $x,y\in V^n$, $x\prec y$ if and only if $x=Dy$
for some doubly stochastic matrix $D$. Here a matrix $D$ is called
doubly stochastic if it is positivity preserving and every row and
column sums to 1. That is,
$$\forall i,j : D_{ij}\geq 0 {\rm \ \ \ and \ \ \ }
\forall j : \sum_i D_{ij} = \sum_i D_{ji} =1.$$ Unfortunately, to
the authors' knowledge, super-majorization does not have such a
correspondence. In order to make use of the known results about
majorization, we must connect probabilistic entanglement
transformation and majorization. The following lemma is just for
this purpose.

\begin{Lemma}\label{lemma:ylambda}
 For $x, y\in V^n$ and $0\leq \lambda \leq 1$, $x\in S^\lambda(y)$ if and only if $x\in
S(y_\lambda)$, where
\begin{equation}\label{ylambda}
y_\lambda=(1-\lambda E_2(y),\lambda y^\downarrow_2,\dots, \lambda
y^\downarrow_n).
\end{equation}
That is, $S^\lambda(y)=S(y_\lambda)$.
\end{Lemma}
\begin{proof} It follows directly from the definitions and we omit the proof here.
\end{proof}

 \vspace{1em}

In the sequel, we expand the notation $y_\lambda$ in
Eq.(\ref{ylambda}) to any probability vectors. Another equivalent
expression of this lemma is that $x\in S^\lambda(y)$ if and only
if $x\prec y_\lambda$. In this paper, we will switch between these
two expressions from time to time for convenience. This simple
lemma is quite useful because it establishes a relationship
between probabilistic transformations and deterministic ones,
while the latter have been well researched.

We know that $S(y)$ is just the convex hull of all vectors which
may be obtained by permutating the components of $y$. A direct
application of the above lemma is a similar description of the
generalized set $S^\lambda(y)$, as the following theorem shows.

\begin{Theorem}\label{theorem:ylambda}
 For all $y\in V^n$ and $0\leq \lambda\leq 1$,
$S^\lambda(y)$ is compact and convex. Furthermore, $S^\lambda(y)$
is the convex hull of the vectors in the following set
$$\{Py_\lambda:\ P\ \ {\rm is\ any\ } n {\rm \ dimesional \ permutation}\}.$$
\end{Theorem}
\begin{proof} Direct from Lemma \ref{lemma:ylambda} and the known
structure of $S(y_\lambda)$.
\end{proof}

\vspace{1em}

The next theorem shows that the set $S^\lambda(y)$, as a function
of $\lambda$, is monotonic, and the intersection of all
$S^\lambda(y)$, $0<\lambda<1$, gives rise to $S(y)$.

\begin{Theorem} Suppose $y\in V^n$ and $0\leq \lambda_1\leq
\lambda_2\leq 1$. Then $S^{\lambda_2}(y)\subseteq
S^{\lambda_1}(y)$. Furthermore, we have that
\begin{equation}\label{lim}
S(y)=\bigcap_{0< \lambda< 1}S^\lambda(y).
\end{equation}
\end{Theorem}
\begin{proof} The monotonicity of $S^\lambda(y)$ is obvious from
the definition. So $$S(y)\subseteq \bigcap_{0< \lambda<
1}S^\lambda(y)$$ holds. To show $$S(y)\supseteq \bigcap_{0<
\lambda< 1}S^\lambda(y),$$ suppose $x\in \bigcap_{0< \lambda<
1}S^\lambda(y)$. It follows that $x\prec y_\lambda$, or
equivalently,
$$E_l(x)\geq E_l(y_\lambda)=\lambda E_l(y)$$
for all $1< l\leq n$ and $0< \lambda< 1$. When $\lambda$ tends to
1, we have $E_l(x)\geq E_l(y)$ for all $1< l\leq n$. Thus $x\in
S(y)$.  This completes the proof.\end{proof}

\vspace{1em}

From the monotonicity of $S^\lambda(y)$ as a function of
$\lambda$, we can define the notions of limit as follows. We call
$S^\lambda(y)$ left continuous at $\lambda$ if for all
nondecreasing sequences $\{\lambda_i:i=1,2,\dots\}$,
$\lim_{i\rightarrow \infty} \lambda_i=\lambda$ implies that
$$\bigcap_{i=1}^\infty S^{\lambda_i}(y) = S^\lambda(y).$$
While $S^\lambda(y)$ is said to be right continuous at $\lambda$
if for all nonincreasing sequences $\{\lambda_i:i=1,2,\dots\}$,
$\lim_{i\rightarrow \infty} \lambda_i=\lambda$  implies that
$$\bigcup_{i=1}^\infty S^{\lambda_i}(y) = S^\lambda(y).$$
Furthermore, $S^\lambda(y)$ is continuous at $\lambda$ if it is
both left continuous and right continuous. Having these notions,
we are able to present the following theorem.

\begin{Theorem} For all $y\in V^n$, $S^\lambda(y)$ is continuous
at any $\lambda$ when $0<\lambda<1$. It is also right continuous
at 0 and left continuous at 1.
\end{Theorem}
\begin{proof}
Easy to check from the definitions. \end{proof}

\vspace{1em}

We have examined thoroughly probabilistic transformations without
catalysis; in the following sections, we will consider
catalyst-assisted ones. At the end of this section, we introduce
some lemmas that are useful for later discussion.

\begin{Lemma}\label{lemma:concave} Given $y\in V^n$, the function $P(x\rightarrow y)$
is concave in $x$.
\end{Lemma}
\begin{proof} For all $x,x'\in V^n$ and $t\in [0,1]$, we have
\[
\begin{array}{rl}
P(tx+(1-t)x'\rightarrow y)&=\min_l\frac{\displaystyle E_l(tx+(1-t)x')}{\displaystyle E_l(y)}\\
\\
&\geq \min_l\frac{\displaystyle E_l(tx)+E_l((1-t)x')}{\displaystyle E_l(y)} \\
\\
&\geq tP(x\rightarrow y)+(1-t)P(x'\rightarrow y).
\end{array}
\]
So $P(x\rightarrow y)$ is concave in $x$. \end{proof}

\vspace{1em}

The next two lemmas consider the properties of the maximal
transformation probability $P(x\rightarrow y)$ under the
operations of direct summation and tensor product on its
parameters $x$ and $y$.

\begin{Lemma}  For all $x,y\in V^n$ and
$x',y'\in V^m$,
$$P(x\oplus x' \rightarrow y\oplus y') \geq \min\{P(x \rightarrow y), P(x'\rightarrow y')\},$$
where $\oplus$ means direct summation. In particular, $P(x\oplus c
\rightarrow y\oplus c) \geq P(x \rightarrow y)$ for all $c$.
\end{Lemma}
\begin{proof}
By Vidal's formula for the probability of entanglement
transformation, there exists an index $l$ such that $1\leq l\leq
n+m$ and
$$P(x\oplus x' \rightarrow y\oplus y')  = \frac{E_l(x\oplus
x')}{E_l(y\oplus y')}.$$ We assume that $E_l(x\oplus
x')=E_{l_x}(x) + E_{l_{x'}}(x')$ for some $l_x\leq n$ and
$l_{x'}\leq m$. Notice that $E_l(y\oplus y')\leq E_{l_x}(y) +
E_{l_{x'}}(y')$ by definition. It follows that
\[
\begin{array}{rl}
P(x\oplus x' \rightarrow y\oplus y') &\geq \frac{\displaystyle
E_{l_x}(x) +
E_{l_{x'}}(x')}{\displaystyle E_{l_x}(y) + E_{l_{x'}}(y')}\\
\\
&\geq \min\{\frac{\displaystyle E_{l_x}(x)}{\displaystyle
E_{l_x}(y)}, \frac{\displaystyle E_{l_{x'}}(x')}{\displaystyle
E_{l_{x'}}(y')}\},
\end{array}
\]
where the second inequality follows from the following fact
$$\frac{a+b}{c+d}\geq \frac{b}{d} \Leftrightarrow
\frac{a}{c}\geq\frac{b}{d}\ \ \ {\rm \ for\ any\ }a,b,c,d\geq 0.
$$
Thus $P(x\oplus x' \rightarrow y\oplus y') \geq \min\{P(x
\rightarrow y), P(x'\rightarrow y')\}$. \end{proof}

\vspace{1em}

\begin{Lemma}\label{Potimes} For all $x,y\in V^n$ and $x',y'\in V^m$,
$$P(x\otimes x'\rightarrow y\otimes y')\geq P(x\rightarrow
y)P(x'\rightarrow y'),$$ where $\otimes$ means tensor product. In
particular, $P(x\otimes c\rightarrow y\otimes c)\geq
P(x\rightarrow y)$ for all $c$.

\end{Lemma}

\begin{proof} This result is obvious from the physical meaning of
$P(x\rightarrow y)$ since the way that separately transforms $x$
into $y$ and $x'$ into $y'$ gives an implementation of
transforming $x\otimes x'$ to $y\otimes y'$. The probability of
success is the multiplication of the probabilities of those two
transformations. That means $P(x\otimes x'\rightarrow y\otimes
y')\geq P(x\rightarrow y)P(x'\rightarrow y').$

We can, however, give a simple pure mathematical proof as follows.
Without loss of generality, we assume that the components of
$x,x',y$, and $y'$ are nonincreasingly arranged, respectively. For
an arbitrarily fixed integer $l$ satisfying $1\leq l\leq nm$, let
$r_i$ be the smallest index of the components of $x'$ in summands
of $E_l(x\otimes  x')$ that have the form $x_i x'_j$, where $1\leq
i\leq n$. That is,
\begin{equation}\label{equ:ri}
r_i=\min\{j:\ x_i x'_j \geq (x\otimes x')^\downarrow_l\}.
\end{equation}
In case of repeated values of components of $x\otimes x'$, we
regard the terms with smaller $i$ to be included in the sum first.
If the set in the right hand side of Eq.(\ref{equ:ri}) is empty
for some $i$ (that is, any term having the form $x_i x'_j$, $1\leq
j\leq m$, does not occur), then let $r_i=m+1$. With these
notations, we can arrange the summands of $E_l(x\otimes x')$ as
$$E_l(x\otimes x')=\sum_{i=1}^n x_i\sum_{j=r_i}^m x'_j.$$
By the definition of $P(x'\rightarrow y')$, we have
$\sum_{j=r_i}^m x'_j \geq  P(x'\rightarrow y')\sum_{j=r_i}^m y'_j$
for all $r_i$. Thus
$$E_l(x\otimes x')\geq P(x'\rightarrow
y') \sum_{i=1}^n x_i\sum_{j=r_i}^m y'_i.$$ Now we rearrange the
summands of $\sum_{i=1}^n x_i\sum_{j=r_i}^m y'_i$ such that
$$\sum_{i=1}^n x_i\sum_{j=r_i}^m y'_i=\sum_{j=1}^m y'_i\sum_{i=t_j}^n x_i$$
for some $1\leq t_1,\dots,t_m\leq n+1$. By the definition of
$P(x\rightarrow y)$, we have $$\sum_{i=t_j}^n x_i \geq
P(x\rightarrow y)\sum_{i=t_j}^n y_i$$ for all $t_j$. Thus
\[
\begin{array}{rl}
E_l(x\otimes x')&\geq P(x'\rightarrow y')P(x\rightarrow
y)\sum\limits_{j=1}^m y'_i\sum\limits_{i=t_j}^n y_i\\
\\
&\geq P(x'\rightarrow y')P(x\rightarrow y)E_l(y\otimes y'),
\end{array}
\]
and $P(x\otimes x'\rightarrow y\otimes y')\geq P(x\rightarrow
y)P(x'\rightarrow y')$ from the arbitrariness of $l$. \end{proof}

\section {Catalyst-assisted Probabilistic
Transformation}

The aim of this section is to consider the case of entanglement
transformations with the aid of catalysis. First, we present a
necessary and sufficient condition for a given probability vector
to serve as a partial catalyst for a certain probabilistic
transformation.

Without loss of generality, we concentrate on catalysts with
nonzero components, since for any probability vector $c$, $c$ and
$c\oplus 0$ have the same catalysis power in the sense that in any
situation, if one serves as a partial catalyst for some
transformation, so does the other for the same transformation.

\begin{Theorem}\label{theorem:PCon} Suppose $x$ and $y$ are two
nonincreasingly arranged $n$-dimensional probability vectors, and
$P(x\rightarrow y)<\min\{x_n/y_n,1\}$. Let
$$L=\{l:1< l< n {\ \ \rm and\ \ \ }
P(x\rightarrow y)=\frac{E_l(x)}{E_l(y)}\}.$$ Then a
nonincreasingly arranged $k$-dimensional probability vector $c$
serves as a partial catalyst for the transformation from $x$ to
$y$, that is,
$$P(x\otimes c\rightarrow y\otimes c)>P(x\rightarrow y),$$
if and only if for all $r_1,r_2,\dots,r_k\in L\cup \{n+1\}$
satisfying  $r_1\geq \dots\geq r_k\not=n+1$, there exist $i$ and
$j$, $1\leq j<i\leq k$, such that
\begin{equation}\label{PCon}
\frac{c_i}{c_j} < \frac{y_{r_j}}{y_{r_i-1}}
\mbox{\hspace{2em}or\hspace{2em}}
\frac{c_i}{c_j}>\frac{y_{r_j-1}}{y_{r_i}}.
\end{equation}

Here, in order to avoid a too complicated statement, we ignore
some extreme cases of Eq.(\ref{PCon}); the condition (\ref{PCon})
should be understood in the following way: whenever one of the two
components of the disjunction in Eq.(\ref{PCon}) contains the
meaningless term $y_{n+1}$, it is considered to be violated
automatically, and the other component is then required.
\end{Theorem}

\begin{proof}
We will prove the theorem by showing that $c$ cannot serve as a
partial catalyst for transforming $x$ into $y$, that is,
$$P(x\otimes c\rightarrow y\otimes c)=P(x\rightarrow y),$$
if and only if there exist $r_1,r_2,\dots,r_k\in L\cup \{n+1\}$
satisfying $r_1\geq \dots\geq r_k\not=n+1$, such that for all
$1\leq j<i\leq k$,
\begin{equation}\label{conds}
\frac{y_{r_j}}{y_{r_i-1}}\leq \frac{c_i}{c_j}\leq
\frac{y_{r_j-1}}{y_{r_i}},
\end{equation}
where any constraint containing the meaningless term $y_{n+1}$ is
considered to be satisfied automatically.

Notice that for all $l$, $1<l\leq nk$, we can arrange the summands
of $E_l(x\otimes c)$ such that
$$E_l(x\otimes c)=\sum_{j=1}^k c_j\sum_{i=r_j}^n x_i,$$
where $1\leq r_j\leq n+1$. The case $r_j=n+1$ means that any term
having the form $c_jx_i$, $1\leq i\leq n$, does not occur. Without
loss of generality, we regard terms with smaller $j$ to be
included in the sum first in case of repeated values of components
of $x\otimes c$. This assumption guarantees that $r_1\geq
\dots\geq r_k$. Furthermore, we exclude the possibility of
$r_1=\dots=r_k=n+1$ from $\sum_j r_j=l$ and $1< l\leq nk$.

From the definition of $E_l(y\otimes c)$ and $P(x\rightarrow y)$,
the following inequalities are easy to check:
\[
\begin{array}{rl}
\frac{\displaystyle E_l(x\otimes c)}{\displaystyle E_l(y\otimes
c)}&\geq \frac{\displaystyle \sum_{j=1}^k c_j\sum_{i=r_j}^n x_i}{\displaystyle \sum_{j=1}^k c_j\sum_{i=r_j}^n y_i}\\
\\
& \geq \frac{\displaystyle P(x\rightarrow y)(\sum_{j=1}^k c_j\sum_{i=r_j}^n y_i)}{\displaystyle \sum_{j=1}^k c_j\sum_{i=r_j}^n y_i}\\
\\
&=P(x\rightarrow y).
\end{array}
\]
The first equality holds if and only if $E_l(y\otimes
c)=\sum_{j=1}^k c_j\sum_{i=r_j}^n y_i$, while the second equality
holds if and only if every $r_j$ is in $L\cup \{n+1\}$ .
Consequently, we see that $P(x\otimes c\rightarrow y\otimes
c)=P(x\rightarrow y)$ if and only if there exist
$r_1,r_2,\dots,r_k\in L\cup \{n+1\}$ such that
\begin{equation}\label{condx}
E_l(x\otimes c)= \sum_{j=1}^k c_j\sum_{i=r_j}^n x_i
\end{equation}
and
\begin{equation}\label{condy}
E_l(y\otimes c)= \sum_{j=1}^k c_j\sum_{i=r_j}^n y_i
\end{equation}
for some $1<l\leq nk$.

In what follows, we derive the conditions presented in
Eq.(\ref{conds}) from Eqs.(\ref{condx}) and (\ref{condy}). In
fact, Eq.(\ref{condy}) means that $\max_{1\leq i\leq
k}\{y_{r_i}c_i\}\leq \min_{1\leq i\leq k}\{y_{r_i-1}c_i\}$, or
equivalently,
\begin{equation}\label{condy1}
\frac{y_{r_j}}{y_{r_i-1}}\leq \frac{c_i}{c_j}\leq
\frac{y_{r_j-1}}{y_{r_i}}
\end{equation}
for all $i,j=1,2,\dots,k$ and $i>j$. The special case of $r_i=n+1$
or $r_j=n+1$ can be included in Eq.(\ref{condy1}) by simply
assuming that the constraints in Eq.(\ref{condy1}) containing the
meaningless term $y_{n+1}$ are automatically satisfied.
Analogously, we can show that Eq.(\ref{condx}) is equivalent to
\begin{equation}\label{condx1}
\frac{x_{r_j}}{x_{r_i-1}}\leq \frac{c_i}{c_j}\leq
\frac{x_{r_j-1}}{x_{r_i}}
\end{equation}
for all $i>j$. Notice that for all $r_i, r_j\in L$,
$$\frac{x_{r_j}}{y_{r_j}}\leq P(x\rightarrow y)\leq
\frac{x_{r_i-1}}{y_{r_i-1}}.$$ It follows that the constraints in
Eq.(\ref{condx1}) can be derived from those in Eq.(\ref{condy1}).
That completes our proof. \end{proof}

Intuitively, if we decompose $x\otimes c$ and $y\otimes c$ as
$$x\otimes c=c_1x\oplus\dots \oplus c_kx$$  and $$y\otimes c=c_1y\oplus\dots\oplus
c_ky,$$ respectively, then when the conditions in
Eq.(\ref{condy1}) are satisfied for some $r_1,r_2,\dots,r_k\in
L\cup \{n+1\}$, we have $c_ix_{r_i}\leq c_jx_{r_j-1}$ and
$c_jx_{r_j}\leq c_ix_{r_i-1}$ for all $1\leq i,j\leq n$. So the
smallest $k(n+1)-\sum_{i=1}^k r_i$ components of $x\otimes c$ are
exactly the components of the form $x_ic_j$, where $1\leq j\leq k$
and $r_j\leq i\leq n$. A similar argument holds for $y\otimes c$.
It follows that when we take $l=\sum_{i=1}^k r_i -k+1$, then
$E_l(x\otimes c)=P(x\rightarrow y)E_l(y\otimes c)$ and thus
$P(x\otimes c\rightarrow y\otimes c)=P(x\rightarrow y)$.

To our surprise, the constraints presented in Eq.(\ref{PCon}) for
the probability vector $c$ to serve as a partial catalyst for
transforming $x$ into $y$ are almost irrelevant to $x$. The only
effect of $x$ is to determine the index set $L$.

\begin{Coro}\label{cor:cor3} Let $x,y,c$, and $L$ be as in the above theorem.
If $$P(x\otimes c\rightarrow y\otimes c)>P(x\rightarrow y),$$ then
\begin{equation}\label{cor3} \frac{c_k}{c_{k-1}}>
\max\{\frac{y_n}{y_{l}}:l\in L\}.
\end{equation}
 \end{Coro}

\begin{proof}
For any $l\in L$, take $r_1=\dots=r_{k-1}=n+1$ and $r_k=l$.
Noticing that the constraints having the term $y_{n+1}$ are
violated automatically, we can reduce Eq.(\ref{PCon}) to the
condition that
\begin{equation}\label{l0}
\frac{c_k}{c_j}> \frac{y_n}{y_{l}}
\end{equation} for some $j<k$. So
$$\frac{c_k}{c_{k-1}}\geq \frac{c_k}{c_j}> \frac{y_n}{y_{l}},$$
and the corollary holds from the arbitrariness of $l$.
\end{proof}

\vspace{1em}

Intuitively, Corollary \ref{cor:cor3} shows that the difference
between the smallest two components of a partial catalyst cannot
be too large. The following corollary, on the other hand, shows
that the difference between the smallest and the largest
components of a partial catalyst cannot be too small. Recall that
a uniform state is a state which has equal nonzero Schmidt
coefficients.

\begin{Coro}
Let $x,y,c$, and $L$ be as in the above theorem. If $$P(x\otimes
c\rightarrow y\otimes c)>P(x\rightarrow y),$$ then

\begin{equation}\label{equ:Coro4}
\frac{c_k}{c_1}< \min\{\frac{y_l}{y_{l-1}}:l\in L\}.
\end{equation} In particular, any uniform state cannot
serve as a partial catalyst for any probabilistic transformation.
\end{Coro}

\begin{proof} For any $l\in L$, let $r_1=\dots=r_k=l$. Then the
conditions in Eq.(\ref{PCon}) become
\begin{equation}\label{equ:uniform}
\frac{c_i}{c_j}< \frac{y_l}{y_{l-1}}
\end{equation}
for some $i>j$. Noticing that $\frac{c_k}{c_1}\leq
\frac{c_i}{c_j}$ for all $1\leq j<i\leq k$ and $l$ is taken
arbitrarily from $L$, we complete the proof.
\end{proof}

 \vspace{1em}

A special and perhaps more interesting case of Theorem
\ref{theorem:PCon} is when $L$ has only one element, that is,
$L=\{l\}$ for some $1<l< n$. In this case, we have the following
corollary.

\begin{Coro}\label{Cor: 2dim} Let $x,y$, and $L$ be as in Theorem \ref{theorem:PCon}. If
$L=\{l\}$ for some $1<l< n$, and
\begin{equation}\label{existc}
\frac{y_{n}}{y_l} < \frac{y_l}{y_{l-1}},
\end{equation}
 then any nonincreasingly arranged $k$-dimensional probability vector $c$ with
\begin{equation}\label{cinte}
\frac{y_{n}}{y_l} <\frac{c_k}{c_t}< \frac{y_l}{y_{l-1}}
\end{equation}
for some $t<k$ serves as a partial catalyst for transforming $x$
into $y$.
\end{Coro}
\begin{proof} Since $L=\{l\}$ contains one element, any
choice of $r_1,r_2,\dots,r_k\in L\cup \{n+1\}$ satisfying $r_1\geq
\dots\geq r_k\not=n+1$ has the form
$$r_1=\dots=r_h=n+1,\ \  r_{h+1}=\dots=r_k=l$$
for some $0\leq h<k$. Take $i=k$ and $j=t$. In what follows we
show that under this choice of $i,j$, Eq.(\ref{cinte}) implies
Eq.(\ref{PCon}) in Theorem \ref{theorem:PCon} and thus $c$ is a
partial catalyst for transforming $x$ into $y$.

In fact, if $t\leq h$, then $r_i=l$ and $r_j=n+1$. Thus
$$\frac{y_{r_j-1}}{y_{r_i}}
=\frac{y_{n}}{y_l} <\frac{c_k}{c_t} =\frac{c_i}{c_j}.$$ Similarly,
if $t> h$ then $r_i=r_j=l$ and
$$\frac{y_{r_j}}{y_{r_i-1}}
=\frac{y_{l}}{y_{l-1}}
>\frac{c_k}{c_t} =\frac{c_i}{c_j}.$$ Thus the conditions in
Eq.(\ref{PCon}) holds and that completes our proof.
\end{proof}

\vspace{1em}

When 2-dimensional catalysts are considered, Eq.(\ref{cinte})
reduces to
\begin{equation}\label{cinte2}
\frac{y_{n}}{y_l} <\frac{c_2}{c_1}< \frac{y_l}{y_{l-1}}.
\end{equation}
Furthermore, it is easy to check that Eq.(\ref{cinte2}) is indeed
a necessary and sufficient condition for a 2-dimensional
probability vetor $c$ to be a partial catalyst. Thus
Eq.(\ref{existc}) also becomes a necessary and sufficient one to
guarantee the transformation from $x$ to $y$ has 2-dimensional
partial catalysts.

From Theorem \ref{theorem:PCon} we can derive a necessary and
sufficient condition for when a given probabilistic transformation
has partial catalysts.

\begin{Theorem}\label{Theorem:catainc}
 Suppose $x,y\in V^n$ and the components of
$x$ and $y$ are nonincreasingly arranged, respectively. Then the
probabilistic transformation from $x$ to $y$ has partial catalysts
if and only if
$$P(x\rightarrow y)<\min\{x_n/y_n,1\}.$$
\end{Theorem}

\begin{proof} The `only if' part is easy and we omit the details
here. The proof of `if' part is as follows.

We abbreviate $P(x\rightarrow y)$ to $P$ for simplicity in this
proof. Let $l_{min}$ and $l_{max}$ be the minimal element and
maximal element in $L$, respectively. Then $y_n<y_{l_{max}}$ since
otherwise
$$\frac{E_{l_{max}}(x)}{E_{l_{max}}(y)}\geq
\frac{(n-l_{max}+1)x_n}{(n-l_{max}+1)y_n} = \frac{x_n}{y_n}
>P,$$ which contradicts the assumption that $l_{max}\in L$. Now
let $\alpha$ be a real number such that $y_n/y_{l_{max}}<\alpha<1$
and $k$ a positive integer such that
$\alpha^{k-1}<y_{l_{max}}/y_{l_{min}-1}$. In what follows, we
prove that $$c=(1,\alpha,\alpha^2,\dots,\alpha^{k-1})$$ will serve
as a partial catalyst for the transformation from $x$ to $y$, that
is,
$$P(x\otimes c\rightarrow y\otimes c)>P(x\rightarrow y).$$
Here we omit the normalization of $c$ for simplicity.

For all $r_1,r_2,\dots,r_k\in L\cup \{n+1\}$ satisfying $r_1\geq
\dots\geq r_k\not=n+1$. There are two cases to consider.

Case 1. $r_1\in L$. In this case, let $i=k$ and $j=1$. Then
$$\frac{c_i}{c_j}= \alpha^{k-1}<\frac{y_{l_{max}}}{y_{l_{min}-1}}
\leq \frac{y_{r_1}}{y_{r_{k}-1}}=\frac{y_{r_j}}{y_{r_i-1}}.$$

Case 2. $r_1=n+1$. In this case, denote by $m$, $1\leq m\leq k-1$,
the (unique) integer such that $r_m=n+1$ but $r_{m+1}\in L$. Now
let $i=m+1$ and $j=m$. Then
$$\frac{c_i}{c_j}=\alpha>\frac{y_n}{y_{l_{max}}}\geq
\frac{y_n}{y_{r_{m+1}}}=\frac{y_{r_j-1}}{y_{r_i}}.$$

In either case, the constraints in Eq.(\ref{PCon}) are satisfied.
So from Theorem \ref{theorem:PCon} we know that the present
theorem holds.
\end{proof}

\vspace{1em} To illustrate the utility of Theorem
\ref{Theorem:catainc}, let us see an example from \cite{JP99} (it
has been presented in Introduction). Let $x=(0.6,0.2,0.2)$ and
$y=(0.5,0.4,0.1)$. We show how to construct a partial catalyst by
the above theorem. It is easy to check that $L=\{2\}$ and
$$\frac{y_3}{y_2}=0.25<0.8=\frac{y_2}{y_1}.$$
So we can take $k=2$ and any two dimensional nonnormalized vector
$c=(1,\alpha)$ for $0.25<\alpha<0.8$ serves as a partial catalyst
for transforming $x$ into $y$. Furthermore, from the remark behind
Corollary \ref{Cor: 2dim}, $\{c=(1,\alpha):0.25<\alpha<0.8\}$ is
exactly the set of all two dimensional partial catalysts for this
transformation.

Suppose now $x$ is just as above while $y=(0.5, 0.3, 0.2)$. Then
$L=\{2\}$ but $y_3/y_2=2/3>y_2/y_1=0.6$. So any two-dimensional
state cannot serve as a partial catalyst for the probabilistic
transformation from $x$ to $y$. We can, however, construct a
higher dimensional partial catalyst from Theorem
\ref{Theorem:catainc} as follows. First, take a real number
$\alpha>y_3/y_2=2/3$. In order not to make $k$ too large, we
should take $\alpha$ as small as possible. For example,
$\alpha=0.67$. Then from the constraint $\alpha^{k-1}<y_2/y_1=0.6$
in the theorem, we have $k\geq 3$. Thus the state
$c=(1,\alpha,\alpha^2)$ can serve as a partial catalyst for
transforming $x$ into $y$.

It is worth noting that the catalyst $c$ presented in the proof of
the above theorem can be replaced by a sequence of 2-dimensional
vectors. To see this, notice that the only constraint on the
dimension $k$ of the catalyst $c$ is
$\alpha^{k-1}<y_{l_{max}}/y_{l_{min}-1}$, that is, for all
sufficiently large $k$,
$$c=(1,\alpha,\alpha^2,\dots,\alpha^{k-1})$$ is an appropriate
partial catalyst which can increase the maximal probability of
transforming $x$ into $y$. In particular, take $k=2^{m+1}-1$ for
some positive integer $m$. From the simple fact that
$$(1,\alpha,\alpha^2,\dots,\alpha^{2^{m+1}-1})=(1,\alpha)\otimes (1,\alpha^2) \otimes \dots \otimes
(1,\alpha^{2^m}),$$ the effect of the catalyst in the left hand
side can be implemented by the sequence of 2-dimensional catalysts
listed in the right hand side. From this observation, a potential
`catalyst bank' need only prepare sufficiently many 2-dimensional
catalysts in order to help probabilistic transformation.

We state the arguments above as the following theorem.

\begin{Theorem}\label{theorem:2diminc}
The set $V^2$ constitutes a complete set of partial catalysts for
all probabilistic entanglement transformations. That is, for all
positive $n$ and $x,y\in V^n$, if $P(x\otimes c\rightarrow
y\otimes c)>P(x\rightarrow y)$ for some $c$, then there exists a
sequence of probability vectors $c_1,\dots,c_m\in V^2$ such that
$$P(x\otimes c_1\otimes\dots \otimes c_m\rightarrow y\otimes c_1\otimes\dots
\otimes c_m)>P(x\rightarrow y).$$
\end{Theorem}

We have presented a necessary and sufficient condition under which
a given entanglement transformation has partial catalysts.
Furthermore, the proof process constructs real catalysts. The
constructed catalysts are, however, not very economical in the
sense that they are usually not with the minimal dimension among
all the probability vectors which can serve as partial catalysts
for this transformation. In what follows, we show how to construct
economical ones.

First, from Theorem \ref{theorem:PCon}, we can design an efficient
algorithm to decide whether a probabilistic transformation has
partial catalysts with a given dimension.

\begin{Theorem}\label{theorem:algorithm} Suppose $x,y$ are two $n$-dimensional
probability vectors and $P(x\rightarrow
y)<\min\{x^\downarrow_n/y^\downarrow_n,1\}$. Let $k$ be a given
positive integer. Then the problem of whether there exists a
$k$-dimensional partial catalyst $c$ for transforming $x$ into $y$
can be decided in polynomial time about $n$.
\end{Theorem} \begin{proof} Without loss of generality, we assume
that the components of $x$, $y$, and $c$ are respectively arranged
nonincreasingly. Notice that from the proof of Theorem
\ref{theorem:PCon} (see Eq.(\ref{condy1})), the necessary and
sufficient condition under which $c$ can increase the maximal
probability of transforming $x$ into $y$ can be reexpressed as,
for all $r_1,r_2,\dots,r_k\in L\cup \{n+1\}$ satisfying $r_1\geq
\dots\geq r_k\not=n+1$,
$$
\max_{1\leq i\leq k} \{y_{r_i}c_i\} > \min_{1\leq i\leq k}
\{y_{r_i-1}c_i\}.
$$
This condition leads to the following algorithm to decide whether
a $k$-dimensional partial catalyst exists for transforming $x$
into $y$:

1. Calculate $P(x\rightarrow y)$ and determine the set $L=\{l:1<
l< n {\ \ \rm and\ \ \ } P(x\rightarrow
y)=\frac{E_l(x)}{E_l(y)}\}.$

2. For all $k$ positive integers $r_1,r_2,\dots,r_k$ chosen from
$L\cup \{n+1\}$, if $r_1\geq \dots\geq r_k\not=n+1$, then solve
the following inequality about $c_1,\dots,c_k$:
\begin{equation}\label{sol}
\max_{1\leq i\leq k} \{y_{r_i}c_i\} > \min_{1\leq i\leq k}
\{y_{r_i-1}c_i\}.
\end{equation}
Then there exists a $k$-dimensional partial catalyst $c$ if and
only if the intersection of the solution areas of Eqs. in
(\ref{sol}) is not empty when $r_1,r_2,\dots,r_k$ range over
$L\cup \{n+1\}$ but satisfy $r_1\geq \dots\geq r_k\not=n+1$.
Notice that the solution area of Eq.(\ref{sol}) for a given
sequence $r_1,r_2,\dots,r_k$ is just the union of those of the
following $k^2$ inequalities:
$$y_{r_i}c_i > y_{r_j-1}c_j,$$ which can be solved in polynomial time of $k$.
Furthermore, the number of choices of $r_1,r_2,\dots,r_k$ is less
than $(\#L+1)^k$, where $\#L$ denotes the number of elements in
$L$ and obviously, $\#L<n-1$. So the algorithm presented above
runs in $O(k)(\#L+1)^k=O(n^k)$ time, which is a polynomial of $n$
when $k$ is treated as a constant. \end{proof}

\vspace{1em}

Notice that in \cite{SD03}, Sun $et$ $al$ have presented a
polynomial time algorithm to decide whether a given entanglement
transformation has $k$-dimensional catalysts, that is,
$k$-dimensional partial catalysts which can increase the maximal
transforming probability to 1. So a little modification of Sun's
algorithm can also be used to determine whether or not a
$k$-dimensional partial catalyst exists. What we would like to
point out is that the complexity of Sun's algorithm is $O(n^{2k})$
while our algorithm presented in Theorem \ref{theorem:algorithm}
is $O((\#L)^k)$. Although they are both exponential of $k$ and in
the worst case $\#L=n-2$, in practice our algorithm is more
efficient since $\#L$ is generally much less than $n$.

Theorem \ref{theorem:algorithm} and Theorem \ref{Theorem:catainc}
together give a method for finding out the most economical
catalysts for a given entanglement transformation as follows.
First, we use Theorem \ref{Theorem:catainc} to decide whether
partial catalysts exist for this transformation and an upper bound
$m$ on the dimensions of the most economical ones can also be
derived. Second, for $k=2,3,\dots,m$ we use the algorithm
presented in Theorem \ref{theorem:algorithm} to decide whether
there exist $k$-dimensional partial catalyst. Moreover, from the
algorithm presented in Theorem \ref{theorem:algorithm}, the most
economical partial catalysts can be constructed explicitly.

\section {Structure of Catalyst-assisted Probabilistic
Transformation}

In this section, we investigate the mathematical structure of
catalyst-assisted probabilistic entanglement transformation. Given
a probability vector $y\in V^n$ and  $0\leq\lambda\leq 1$, denote
by $T^\lambda(y)$ the set of probability vectors which, with the
aid of some catalyst, can be transformed into $y$ with a
probability not less than $\lambda$, that is
$$T^\lambda(y)=\{x\in V^n:P(x\otimes c\rightarrow y\otimes c)\geq
\lambda {\rm\ \ for\ some}$$
 $$ {\rm probability\ vector\ } c\}.$$ The
special case of $\lambda=1$ corresponds to $T(y)$, which is just
the set of all probability vectors that can be transformed
deterministically into $y$ by catalyst-assisted LOCC (for the
definition, we refer to \cite{DK01} or \cite{DF03}). It is easy to
check that the set $T^\lambda(y)$, as a function of $\lambda$, is
monotonic.

Recall that $S^\lambda(y)$ is just equal to $S(y_\lambda)$ from
Lemma \ref{lemma:ylambda}. One may wonder if
$T^\lambda(y)=T(y_\lambda)$, or if there exists a simple
connection between catalyst-assisted probabilistic transformation
and catalyst-assisted deterministic one. If so, then all the known
properties of $T(y_\lambda)$ can be used to give simple proofs of
those of $T^\lambda(y)$. In fact, we can prove the following.
\begin{Lemma} For $x, y\in V^n$, if $x\in T(y_\lambda)$, then
$x\in T^\lambda(y)$. That is, $T(y_\lambda)\subseteq
T^\lambda(y)$. \end{Lemma} \begin{proof} Suppose $x\in
T(y_\lambda)$. By definition, there exists a probability vector
$c$ such that $x\otimes c\prec y_\lambda\otimes c$. Noticing that
$$y_\lambda\otimes c = (1-\lambda E_2(y))c\oplus \lambda
y_2c\oplus\dots\oplus \lambda y_n c \prec (y\otimes c)_\lambda,$$
we have $x\otimes c\prec (y\otimes c)_\lambda$, which implies that
$x\in T^\lambda(y)$. \end{proof}

\vspace{1em}

But unfortunately, $T(y_\lambda)\not= T^\lambda(y)$. Moreover, we
can show informally that for all $z\in V^n$, $T(z)\not=
T^\lambda(y)$ as follows. From Corollary 5 in \cite{DK01}, the
boundary points of $T(z)$ is the set $\{x\in
T(z):x^\downarrow_1=z^\downarrow_1 {\rm \ \ or\ \ }
x^\downarrow_n=z^\downarrow_n\}$, while from Theorem
\ref{theorem:boundary} in this section, the boundary points of
$T^\lambda(y)$ is the set $\{x\in T^\lambda(y):
x^\downarrow_n=\lambda y^\downarrow_n\}$. It is obvious that these
two sets cannot be equal. So $T(z)\not= T^\lambda(y)$. Since
$T^0(y)=V^n$ and $T^1(y)=T(y)$, we assume from now on that
$0<\lambda<1$. The next two lemmas are useful for latter
discussion.

\begin{Lemma}\label{lemma:Soplus} Let $x,y\in V^n$ and $0<\lambda<1$. Suppose $x$ and
$y$ can be decomposed into two parts respectively as
$$x=x'\oplus \lambda z {\rm\ and\ } y=y'\oplus z$$ for some $z$
such that $z^\downarrow_1<y^\downarrow_1$. Let $\mu$ denote the
sum of the components of $z$. Then $x\in S^\lambda(y)$ if and only
if
$$\frac{x'}{1-\mu\lambda} \in S^{\lambda'}(\frac{y'}{1-\mu}),$$
where
\begin{equation}\label{lambda}
\lambda'=\frac{\lambda-\mu\lambda}{1-\mu\lambda}.
\end{equation}
\end{Lemma}
\begin{proof} Without loss of generality, we can assume that
$$y'=(y_{i_1}, y_{i_2}, \dots, y_{i_m}),$$
and the components of $y'$ have been arranged nonincreasingly. If
$x\in S^\lambda(y)$, then $x\prec y_\lambda$ from Lemma
\ref{lemma:ylambda}. Furthermore, we have $y_\lambda=y'' \oplus
\lambda z$ and then
$$\frac{x'}{1-\mu\lambda} \prec \frac{y''}{1-\mu\lambda},$$
where
$$y''=(1-\lambda E_2(y), \lambda y_{i_2}, \dots, \lambda y_{i_m}).$$
A simple calculation shows that
$$ \frac{y''}{1-\mu\lambda}=(\frac{y'}{1-\mu})_{\lambda'},$$ so
the `only if' part of the lemma holds.

The `if' part can be proved by simply retracing the arguments
above. \end{proof}

\begin{Lemma}\label{lemma:Toplus} Let $x,y,\lambda,\mu$, and $\lambda'$ be as in Lemma
\ref{lemma:Soplus}. Then $x\in T^\lambda(y)$ if and only if
$$\frac{x'}{1-\mu\lambda} \in T^{\lambda'}(\frac{y'}{1-\mu}).$$
\end{Lemma}
\begin{proof} Notice that $x\in T^\lambda(y)$ if and only if there exists a
probability vector $c$ such that $x\otimes c\in S^\lambda
(y\otimes c)$. The current lemma is just a simple application of
Lemma \ref{lemma:Soplus}.
\end{proof}

The following theorem shows that some properties of $T(y)$ such as
convexity and containing corresponding no-catalysis case as a
subset are also shared by $T^\lambda(y)$.

\begin{Theorem}\label{theorem:sufficient} For all $y\in V^n$
and $0< \lambda< 1$,

1) $T^\lambda(y)$ is convex;

2) $S^\lambda(y)\subseteq T^\lambda(y)$;

3) suppose $x\in T^\lambda(y)$ and the components of $x$ and $y$
are nonincreasingly arranged respectively. If $x\not=y_\lambda$,
then $x_d/y_d> \lambda$, where $d$ is the maximal index of the
components such that $x_d/y_d\not= \lambda$.
\end{Theorem}
\begin{proof} 1) Suppose $x,x'\in T^\lambda (y)$. By
definition, there exist probability vectors $c$ and $c'$ such that
$P(x\otimes c\rightarrow y\otimes c)\geq \lambda$ and $P(x'\otimes
c'\rightarrow y\otimes c')\geq \lambda$. For all $t$, $0\leq t\leq
1$, we have
\[
\begin{array}{rl}
&P((tx+(1-t)x')\otimes \widetilde{c}\rightarrow y\otimes \widetilde{c})\\
\\
&\geq tP(x\otimes \widetilde{c}\rightarrow y\otimes
\widetilde{c})+(1-t)P(x'\otimes \widetilde{c}\rightarrow y\otimes \widetilde{c})\\
\\
&\geq tP(x\otimes c\rightarrow y\otimes c) + (1-t)P(x'\otimes
c'\rightarrow y\otimes c')\geq \lambda
\end{array}
\]
from Lemma \ref{lemma:concave} and Lemma \ref{Potimes}, where
$\widetilde{c}=c\otimes c'$.

2) It is obvious from Lemma \ref{Potimes}.

3) Suppose $x\in T^\lambda(y)$. Decompose $x$ as $x=x'\oplus x''$
where
$$x'=(x_1,x_2,\dots,x_d) {\ \ \ \rm and\ \ \ }x''=(x_{d+1},\dots,x_n).$$
The vector $y$ is decomposed analogously. From the definition of
$d$, we have $x''=\lambda y''$. Applying Lemma \ref{lemma:Toplus},
we have
\begin{equation}\label{lambdaprime}
\frac{x'}{1-\mu\lambda} \in T^{\lambda'}(\frac{y'}{1-\mu}),
\end{equation}
where $\mu$ is the sum of the components of $y''$ and $\lambda'$
is defined by Eq.(\ref{lambda}). We can then deduce that
$x_d/y_d>\lambda$ since otherwise (notice that
$x_d/y_d\not=\lambda$ by assumption)
\[
\begin{array}{rl}
P(\frac{\displaystyle x'}{\displaystyle 1-\mu\lambda}\otimes
c\rightarrow \frac{\displaystyle y'}{\displaystyle 1-\mu}\otimes
c)&\leq \frac{\displaystyle (1-\mu)x'_d
c_k}{\displaystyle (1-\mu\lambda)y'_d c_k}\\
\\
&=\frac{\displaystyle (1-\mu)x_d}{\displaystyle (1-\mu\lambda)y_d}
<\lambda'
\end{array}
\]
for any probability vector $c$, where $k$ is the dimension of $c$.
This contradicts Eq.(\ref{lambdaprime}).
\end{proof}

Notice that in 3), no constraints on the largest components of $x$
and $y$ are needed for $x\in T^\lambda(y)$, in contrast to the
necessary condition that $x_1\leq y_1$ of $x\in T(y)$. This is due
to the asymmetry of roles of the largest and the smallest
components in determining the maximal transforming probability.

\begin{Lemma}\label{lemma:boundary} Let $y\in V^n$ and $0<\lambda<1$. For
all $x\in S^\lambda(y)$, if $P(x\rightarrow
y)<x^\downarrow_n/y^\downarrow_n$, then $x$ is in the interior of
$T^\lambda(y)$.
\end{Lemma}
\begin{proof} Suppose $x\in S^\lambda(y)$ and $P(x\rightarrow
y)<x^\downarrow_n/y^\downarrow_n$. If $P(x\rightarrow y)=1$, then
$x$ is in the interior of $S^\lambda(y)$ for $\lambda<1$. Thus $x$
is also an interior point of $T^\lambda(y)$ since
$S^\lambda(y)\subseteq T^\lambda(y)$. When $P(x\rightarrow y)<1$,
by Theorem \ref{Theorem:catainc}, there exists a partial catalyst
$c$ such that $P(x\otimes c\rightarrow y\otimes c)>P(x\rightarrow
y)\geq\lambda$. Thus $x$ is an interior point of $T^\lambda(y)$.
\end{proof}

 \begin{Theorem}\label{theorem:boundary} Let $y\in V^n$, $0<\lambda <1$ and $x\in
T^\lambda(y)$. Then $x$ is on the boundary of $T^\lambda(y)$ if
and only if $x^\downarrow_n/y^\downarrow_n=\lambda$.
\end{Theorem}
\begin{proof} Notice first that $x\in T^\lambda(y)$ implies
$x^\downarrow_n/y^\downarrow_n\geq\lambda$ from Theorem
\ref{theorem:sufficient}. Suppose now that
$x^\downarrow_n/y^\downarrow_n=\lambda$. For all $\epsilon>0$,
consider the probability vector
$$x'=(x^\downarrow_1,\dots,x^\downarrow_{n-2},x^\downarrow_{n-1}+\epsilon,x^\downarrow_n-\epsilon).$$
It is easy to check that
$$P(x'\otimes c\rightarrow y\otimes
c)\leq\frac{(x'\otimes c)^\downarrow_{nk}}{(y\otimes
c)^\downarrow_{nk}}=\frac{x^\downarrow_n-\epsilon}{y^\downarrow_n}<
\frac{x^\downarrow_n}{y^\downarrow_n}=\lambda$$ for any
probability vector $c$, where $k$ is the dimension of $c$. Thus
$x'\not\in T^\lambda(y)$. It follows that $x$ is a boundary point
of $T^\lambda(y)$.

Conversely, suppose $x^\downarrow_n/y^\downarrow_n>\lambda$. By
the definition of $x\in T^\lambda(y)$, there exists a probability
vector $c$ such that $P(x\otimes c\rightarrow y\otimes c)\geq
\lambda$. If the inequality holds strictly, then $x$ is of course
in the interior of $T^\lambda(y)$ by the continuum of $P(x\otimes
c\rightarrow y\otimes c)$ on $x$. So we need only consider the
case of $P(x\otimes c\rightarrow y\otimes c)= \lambda$. In this
case we have $$P(x\otimes c\rightarrow y\otimes c)=
\lambda<\frac{x^\downarrow_n}{y^\downarrow_n}=\frac{(x\otimes
c)^\downarrow_{nk}}{(y\otimes c)^\downarrow_{nk}},$$ where $k$ is
the dimension of $c$. Thus $x\otimes c$ is an interior point of
$T^\lambda(y\otimes c)$ from Lemma \ref{lemma:boundary}. On the
other hand, since the function $x\mapsto x\otimes c$ is
continuous, it follows that $x$ is in the interior of the set
$\{x:x\otimes c\in T^\lambda(y\otimes c)\}$, which is just a
subset of $T^\lambda(y)$. So $x$ is in the interior of
$T^\lambda(y)$.
\end{proof}

\vspace{1em}

Recall that the boundary point set of $T(y)$ is $\{x\in
T(y):x^\downarrow_1=y^\downarrow_1 {\rm\ or\
}x^\downarrow_n=y^\downarrow_n\}$. Once again, the asymmetry of
roles of the largest and the smallest components in determining
the maximal transforming probability makes the boundary point set
of $T^\lambda(y)$ different from that of $T(y)$.

The next theorem tells us when catalysis is helpful for
probabilistic transformation with destination state $y$ by giving
a necessary and sufficient condition for
$T^\lambda(y)=S^\lambda(y)$.

\begin{Theorem}\label{theorem:equal} Let $y\in V^n$ be a nonincreasingly arranged
probability vector and $0<\lambda<1$. Then
$T^\lambda(y)=S^\lambda(y)$ if and only if $y_2=y_n$.
\end{Theorem}
\begin{proof} If $y_2=y_n$, then for any
nonincreasingly arranged $x\in V^n$ and any integer $l$ satisfying
$1<l\leq n$,
$$\frac{E_l(x)}{E_l(y)}=\frac{\sum_{i=l}^n x_i}{(n-l+1)y_n}
\geq \frac{(n-l+1)x_n}{(n-l+1)y_n}=\frac{x_n}{y_n}.$$ Thus
$P(x\rightarrow y)=\min\{x_n/y_n,1\}$ and for any vector $c$,
$P(x\otimes c\rightarrow y\otimes c)= P(x\rightarrow y)$. It
follows that $T^\lambda(y)=S^\lambda(y)$.

Conversely, suppose $y_2>y_n$. Let $m>1$ be the maximal index of
the components of $y$ which are not equal to $y_n$, that is,
$y_m>y_{m+1}=\dots=y_n$. Then we have $E_m(y)>(n-m+1)y_n$. Let
$\mu$ be a real number such that $$\lambda<\mu<\min\{
\frac{\lambda E_m(y)}{(n-m+1)y_n},1\},$$ and define a probability
vector
$$x=(y_1+\frac{1-\lambda}{m-1}E_m(y),\dots,y_{m-1}+\frac{1-\lambda}{m-1}E_m(y),$$
$$
\lambda E_m(y)-\mu E_{m+1}(y),\mu y_{m+1},\dots,\mu y_n).$$ It is
a little tedious but very easy to check that the components of $x$
have been nonincreasingly arranged and $P(x\rightarrow
y)=\lambda<\mu=x_n/y_n$. By Theorem \ref{theorem:boundary}, $x$ is
an interior point of $T^\lambda(y)$. That completes our proof that
$S^\lambda(y)\not=T^\lambda(y)$ since $x$ is obviously on the
boundary of $S^\lambda(y)$. \end{proof}

 \vspace{1em}
Compared with the conditions of $T(y)=S(y)$ presented in
\cite{DK01}, our condition in Theorem \ref{theorem:equal} is more
simple and even a little surprising since it depends only on
whether or not $y_2=y_n$ (totally irrelevant to the value of
$\lambda$).

In what follows, we consider the interesting question of whether
or not there exists a bound on the dimension of partial catalysts
that we should consider to help transforming some vector $x$ into
a given probability vector $y$. This is in fact a generalization
of the problem considered for $T(y)$ in \cite{DK01}. We will give
a negative answer to this question by showing that in general
$T^\lambda(y)\not = T_k^\lambda(y)$ for all positive $k$, where
$$T_k^\lambda(y)=\{x\in V^n:P(x\otimes c\rightarrow y\otimes
c)\geq \lambda {\rm\ \ for\ some\ }c\in V^k\}$$ is defined to be
the set of probability vectors which, with the aid of some
$k$-dimensional catalyst vector, can be transformed into $y$ with
the maximal probability not less than $\lambda$.

\begin{Lemma}\label{lemma:tk} For all $y\in V^n$ and any positive integer
$k$, $T_k^\lambda(y)$ is a closed set.
\end{Lemma}
\begin{proof}
Suppose $x^1,x^2,\dots$ is an arbitrary vector sequence in
$T_k^\lambda(y)$ that converges to $x$. By the definition of
$T_k^\lambda(y)$, there exists a catalyst sequence $c^1,c^2,\dots$
in $V^k$ such that $P(x^i\otimes c^i\rightarrow y\otimes c^i)\geq
\lambda$ for $i=1,2,\dots$. Notice that $V^k$ is a compact set in
$R^k$. There exists a convergent subsequence
$c^{i_1},c^{i_2},\dots$ of $c^1,c^2,\dots$ that converges to, say,
$c\in V^k$. Then we can deduce that $P(x\otimes c\rightarrow
y\otimes c)\geq \lambda$ and so $x\in T_k^\lambda(y)$ from the
fact that the function $P(x\otimes c\rightarrow y\otimes c)$ is
continuous on the parameters $x$ and $c$. \end{proof}

\vspace{1em}

Notice that in \cite{DK01}, a similar lemma about $T_k(y)$ was
presented but the proof there was a little complex. The proof
technique of the above lemma can be used to give a simpler one.

\begin{Theorem}\label{theorem:notequal} Let $y\in V^n$ be a nonincreasingly arranged
probability vector, $y_2>y_n$, and $0<\lambda<1$. Then
$T^\lambda(y)\not=T^\lambda_k(y)$ for all positive $k$.
\end{Theorem}
\begin{proof} We will complete the proof by showing that under the assumptions
in this theorem, $T^\lambda(y)$ is not a closed set. Then from
Lemma \ref{lemma:tk}, we have $T^\lambda(y)\not=T^\lambda_k(y)$
for all positive $k$.

Let $e=(1/n,\dots,1/n)$ be the uniform vector in $V^n$. Notice
that $e\in T^\lambda(y)$ and from Theorem \ref{theorem:boundary},
$e$ is also an interior point of $T^\lambda(y)$. Denote by $m$
the maximal index of the components of $y$ which are not equal to
$y_n$, that is,
\begin{equation}\label{y2n}
y_m>y_{m+1}=\dots=y_n.
\end{equation}
Since $y_2>y_n$, we have $1<m<n$. Let $\mu\in(0,\lambda)$ and
define a probability vector
$$x=(y_1+\frac{1-\mu}{m-1}E_m(y),\dots,y_{m-1}+\frac{1-\mu}{m-1}E_m(y),$$
$$
\mu y_m,\dots,\mu y_n).$$ It is easy to check that the components
of $x$ are arranged nonincreasingly and $P(x\rightarrow
y)=x_n/y_n=\mu<\lambda$. So $x\not\in T^\lambda(y)$. If
$T^\lambda(y)$ is closed, then the closed set $\{tx+(1-t)e:0\leq
t\leq 1\}$ must intersect $T^\lambda(y)$ at a boundary point of
$T^\lambda(y)$, say, $x'=t'x+(1-t')e$. Thus $x'_n/y_n=\lambda$
from Theorem \ref{theorem:boundary}. On the other hand, from
Eq.(\ref{y2n}) and noticing that $0<t'<1$, we have
$$\frac{x'_n}{y_n}=t'\mu  + \frac{1-t'}{ny_n}>t'\mu  + \frac{1-t'}{ny_m}
=\frac{x'_m}{y_m},$$ and so
$$\frac{x'_m}{y_m}<\frac{x'_{m+1}}{y_{m+1}}=\dots=\frac{x'_n}{y_n}=\lambda,$$
which contradicts 3) of Theorem \ref{theorem:sufficient}.
\end{proof}

\vspace{1em}

Notice that when $y_2=y_n$, we have $T^\lambda(y)=S^\lambda(y)$
from Theorem \ref{theorem:equal}, and so
$T^\lambda(y)=T^\lambda_k(y)$ for any positive integer $k$ since
$$T^\lambda(y)=S^\lambda(y)\subseteq T^\lambda_k(y)\subseteq T^\lambda(y).$$

In order to describe $T^\lambda(y)$ more deeply, we examine its
extreme points in what follows.

The following lemma describes what kind of perturbations will not
remove a point from $T^\lambda(y)$, even when the point is on the
boundary. This lemma can also be regarded as a generalization of
Corollary 5 in \cite{DK01}.
\begin{Lemma}\label{lemma:Perturbation} Suppose $0<\lambda<1$, and $x,y\in V^n$ be two
nonincreasingly arranged $n$-dimensional probability vectors
satisfying $x\in T^\lambda(y)$ but $x\not =y_\lambda$. Let $d$ be
the maximal index of the components such that $x_d>\lambda y_d$.
Then any sufficiently small perturbation will not remove $x$ from
$T^\lambda(y)$, provided that the perturbation does not affect the
components $x_{d+1},\dots,x_n$. (Notice that if $d=n$, then $x$ is
an interior point of $T^\lambda(y)$ and the result here is just a
simple property of interior points.)
\end{Lemma}
\begin{proof} To begin with, let us decompose $x$ as $x=x'\oplus x''$ where
$$x'=(x_1,x_2,\dots,x_d) {\ \ \ \rm and\ \ \ }x''=(x_{d+1},\dots,x_n).$$
Similarly, $y$ can be decomposed as $y=y'\oplus y''$. Applying
Lemma \ref{lemma:Toplus}, we have
\begin{equation}
x\in T^\lambda(y)\ \ \Leftrightarrow\ \ \widetilde{x} \in
T^{\lambda'}(\widetilde{y}),
\end{equation}
where $$\widetilde{x}= \frac{x'}{1-\mu\lambda} {\rm\ \ and\ \ }
\widetilde{y}=\frac{y'}{1-\mu},$$ and $\mu$ is the sum of the
components of $y''$ and $\lambda'$ is defined by
Eq.(\ref{lambda}). Since $x_d>\lambda y_d$, we have
$$\widetilde{x}_d=\frac{x_d}{1-\mu\lambda}>\lambda' \frac{y_d}{1-\mu}=\lambda' \widetilde{y}_d,$$
and so $\widetilde{x}$ is an interior point of
$T^\lambda(\widetilde{y})$ from Theorem \ref{theorem:boundary}.
Notice that any perturbation to $x$ which does not affect the
components $x_{d+1},\dots,x_n$ is also a perturbation to
$\widetilde{x}$ and vice versa. Therefore the lemma follows.
\end{proof}

\vspace{1em}

Using this lemma, we can easily find out all the extreme points of
$T^\lambda(y)$ for a given $y\in V^n$. It is rather surprising
that $T^\lambda(y)$ and $S^\lambda(y)$ share the same extreme
points, although they satisfy very different properties and also
$S^\lambda(y) \subsetneqq T^\lambda(y)$ in general. Notice that
$T^\lambda(y)$ is not closed in general by the proof of Theorem
\ref{theorem:notequal}, so we cannot determine the whole set
$T^\lambda(y)$ only by its extreme points.

\begin{Theorem} For all  $y\in V^n$,
the set of extreme points of $T^\lambda(y)$ is the same as that of
$S^\lambda(y)$. That is, it is just the set
$$\{Py_\lambda:\ P\ \ {\rm is\ any\ } n {\rm \ dimesional \
permutation}\}.$$
\end{Theorem}
\begin{proof} First, we prove that any extreme point of
$S^\lambda(y)$ is an extreme point of $T^\lambda(y)$. To prove
this, we need only show that $y_\lambda$ is an extreme point of
$T^\lambda(y)$. Suppose the components of $y$ have been arranged
nonincreasingly and
\begin{equation}\label{combination}
y_\lambda=ty'+(1-t)y''
\end{equation}
is a convex combination of $y'$ and $y''$ in $T^\lambda(y)$, where
$0<t<1$. From 3) of Theorem \ref{theorem:sufficient} we have
$$y'_n\geq y'^\downarrow_n\geq \lambda y_n$$ and
$$y''_n\geq y''^\downarrow_n\geq \lambda y_n$$
(notice that $y'$ and $y''$ are not necessarily nonincreasingly
arranged). But from Eq.(\ref{combination}) we have
$ty'_n+(1-t)y''_n=\lambda y_n$. It follows that
\begin{equation}\label{yn}
y'_n=y''_n=\lambda y_n,
\end{equation}
and furthermore, $y'_n$ and $y''_n$ are the smallest components of
$y'$ and $y''$, respectively. Now using 3) of Theorem
\ref{theorem:sufficient} again, we can deduce that
$$y'_{n-1}=y''_{n-1}=\lambda y_{n-1}$$
from Eqs.(\ref{yn}) and (\ref{combination}). And again, $y'_{n-1}$
and $y''_{n-1}$ are the second smallest components of $y'$ and
$y''$, respectively. Repeating the above arguments, we finally
come to the conclusion that
$$y'=y''=y_\lambda.$$
Thus $y_\lambda$ is an extreme point of $T^\lambda(y)$ as claimed.

Now suppose $x$ is an extreme point of $T^\lambda(y)$ and $x$ is
nonincreasingly arranged. If it is not an extreme point of
$S^\lambda(y)$, then $x\not = y_\lambda$ and so there exists a
positive integer $d$, $1< d\leq n$, such that $x_d>\lambda y_d$.
By Lemma \ref{lemma:Perturbation}, we can find a sufficiently
small but positive $\epsilon$ such that $x'\in T^\lambda(y)$ and
$x''\in T^\lambda(y)$, where $$x'=
(x_1,\dots,x_{d-2},x_{d-1}-\epsilon,x_d+\epsilon,x_{d+1},\dots,x_n)$$
and
$$x''=(x_1,\dots,x_{d-2},x_{d-1}+\epsilon,x_d-\epsilon,x_{d+1},
\dots,x_n).$$ Obviously $x=(x'+x'')/2$, which contradicts our
assumption that $x$ is an extreme point of $T^\lambda(y)$. That
completes our proof.\end{proof}

What we would like to point out here is that a similar argument on
$T(y)$ instead of $T^\lambda(y)$ in the above theorem can lead to
a solution to the open problem Nielsen proposed in his lecture
notes on the theory of majorization and its applications in
quantum information theory \cite{MAJ}. More specifically, $T(y)$
has a discrete set, but not a continuum as Nielsen conjectured, of
extreme points (in fact, the extreme points of $T(y)$ are just
$\{Py\ : \mbox{where $P$ is any $n$-dimensional permutation}\}$).

\section{Conclusion and Open problems}

In this paper, we investigate carefully the power of catalysis in
probabilistic entanglement transformations by LOCC. We give a
necessary and sufficient condition for when some appropriately
chosen catalyst can be helpful in increasing the maximal
probability for a given probabilistic transformation. An efficient
algorithm is presented to decide whether partial catalysts with a
given dimension exist for a certain probabilistic transformation,
which leads to a method for constructing the most economical
partial catalysts with minimal dimension. We also study the set of
states that can be transformed by catalyst-assisted LOCC into a
given state with the maximal probability not less than a given
positive number. We prove that this set shares many properties
with the well known set consisting of all vectors being trumped by
the given state. More mathematical structure of catalyst-assisted
probabilistic transformation is also considered.

The emphasis of this paper is to determine when the maximal
transforming probability can be increased in the presence of
partial catalysts and how to construct appropriate ones. The
amount of the probability increased is, however, not considered.
So an open problem and also an important direction for further
study is to determine the maximum of the transforming probability
that can be reached with the aid of partial catalysts. This is
also a generalization of deterministic catalysis since the problem
of the existence of catalysts for deterministic transformations is
equivalent to the problem of the existence of partial catalysts
which can increase the transforming probability to 1 for
probabilistic transformations.

At the end of Section IV, we have determined all the extreme
points of $T^\lambda(y)$. However, we still know little about the
geometric structure of $T^\lambda(y)$. The main reason is that
$T^\lambda(y)$ is not closed in general and so it is not the
convex hull of its extreme points. How to determine the
accumulation points outside $T^\lambda(y)$ is really a hard
problem and remains open. We believe that
$\overline{T}^\lambda(y)$, the closure of $T^\lambda(y)$, has a
continuum of extreme points, just as Nielsen conjectured.

\section*{Acknowledgment}
The authors are grateful to the two referees. Their detailed
comments and suggestions have greatly improved the quality of the
paper.

\end{document}